\documentclass[twocolumn,twoside,slac]{revtex4}
\usepackage{graphicx}
\usepackage{fancyhdr}
\begin{document}

\title{The RooFit toolkit for data modeling}


\author{W. Verkerke}
\affiliation{University of California Santa Barbara, Santa Barbara, CA 93106, USA}
\author{D. Kirkby}
\affiliation{University of California Irvine, Irvine CA 92697, USA}

\begin{abstract}
{\tt RooFit} is a library of C++ classes that facilitate data modeling in
the ROOT environment. Mathematical concepts such as variables,
(probability density) functions and integrals are represented as C++
objects. The package provides a flexible framework for building
complex fit models through classes that mimic math operators, and is
straightforward to extend. For all constructed models {\tt RooFit} provides
a concise yet powerful interface for fitting (binned and unbinned
likelihood, $\chi^2$), plotting and toy Monte Carlo generation as well as
sophisticated tools to manage large scale projects. {\tt RooFit} has matured
into an industrial strength tool capable of running the BABAR
experiment's most complicated fits and is now available to all users
on SourceForge \cite{sourceforge}.
\end{abstract}

\maketitle

\thispagestyle{fancy}


\section{Introduction}

% slide 1 - Purpose

One of the central challenges in performing a physics analysis is to
accurately model the distributions of observable quantities $\vec{x}$
in terms of the physical parameters of interest $\vec{p}$ as well as
other parameters $\vec{q}$ needed to describe detector effects such as
resolution and efficiency. The resulting model consists of a
``probability density function'' (PDF) $F(\vec{x}\,;\vec{p},\vec{q})$
that is normalized over the allowed range of the observables
$\vec{x}$ with respect to the parameters $\vec{p}$ and $\vec{q}$.

Experience in the BaBar experiment has demonstrated that the
development of a suitable model, together with the tools needed to
exploit it, is a frequent bottleneck of a physics analysis. For
example, some analyses initially used binned fits to small samples to
avoid the cost of developing an unbinned fit from scratch. To address
this problem, a general-purpose toolkit for physics analysis modeling
was started in 1999. This project fills a gap in the particle
physicists' tool kit that had not previously been addressed.

A common observation is that once physicists are freed from the
constraints of developing their model from scratch, they often use
many observables simultaneously and introduce large numbers of
parameters in order to optimally use the available data and
available control samples. 

% For example, the showcase BaBar $\sin(2\beta)$ fit,
% implemented by Verkerke in {\ttfamily RooFit}, is five-dimensional and
% involves 45 free parameters.

\section{Overview}

% slide 2 - Implementation
% slide 3 - Desired functionality

The final stages of most particle physics analysis are performed in an
interactive data analysis framework such as PAW\cite{paw} or
ROOT\cite{root}. These applications provide an interactive environment
that is programmable via interpreted macros and have access to a
graphical toolkit designed for visualization of particle physics data.
The {\tt RooFit} toolkit extends the ROOT analysis environment 
% via two shared libraries 
by providing, in addition to basics visualization and
data processing tools, a language to describe data models.  
% {\tt RooFit} does not replace existing functionality in ROOT such as the
% MINUIT minimization package but complements it and interfaces to it. 

The core features of {\tt RooFit} are

\begin{itemize}

\item A natural and self-documenting vocabulary to build a model in
terms of its building blocks (e.g., exponential decay, Argus function,
Gaussian resolution) and how they are assembled (e.g., addition,
composition, convolution). A template is provided for users to add new
building-block PDFs specific to their problem domain.

\item A data description language to specify the observable quantities
being modeled using descriptive titles, units, and any cut
ranges. Various data types are supported including real valued and
discrete valued (e.g. decay mode). Data can be read from ASCII files
or ROOT ntuples.

\item Generic support for fitting any model to a dataset using a
(weighted) unbinned or binned maximum likelihood, or
$\chi^2$ approach

\item Tools for plotting data with correctly calculated errors,
Poisson or binomial, and superimposing correctly normalized
projections of a multidimensional model, or its components.

\item Tools for creating a event samples from any model with Monte Carlo
techniques, with some variables possibly taken from a prototype
dataset, e.g. to more accurately model the statistical fluctuations in
a particular sample.

\item Computational efficiency. Models coded in {\tt RooFit}
should be as fast or faster than hand coded models.  An array of
automated optimization techniques is applied to any model without
explicit need for user support.

\item Bookkeeping tools for configuration management, automated PDF
creation and automation of routine tasks such as goodness-of-fit
studies.

\end{itemize}

\section{Object-Oriented Mathematics}

% slide 4 - Mathematical formulation
% slide 5 - OO representation
% slide 6 - Constructing composite objects
% slide 7 - Bookkeeping

To keep the distance between a physicists' mathematical description of
a data model and its implementation as small as possible, the {\tt RooFit}
interface is styled after the language of mathematics.  The
object-oriented ROOT environment is ideally suited for this approach:
each mathematical object is represented by a C++ software object. 
Table \ref{tab:map} illustrates the correspondence between some basic
mathematical concepts and {\tt RooFit} classes.

\begin{table}[hbt]
\begin{center}
\begin{tabular}{rcl} 
Concept  & Math Symbol & {\tt RooFit} class name \\
\hline
Variable & $x,p$       &  {\tt RooRealVar} \\
Function & $f(\vec{x})$ & {\tt RooAbsReal} \\
PDF      & $F(\vec{x};\vec{p},\vec{q})$ & {\tt RooAbsPdf} \\
Space point & $\vec{x}$  & {\tt RooArgSet} \\
Integral & $~\int_{\vec{x}_{min}}^{\vec{x}_{max}} f(\vec{x}) d\vec{x}~$ & {\tt RooRealIntegal} \\
List of space points & $\vec{x}_k$ & {\tt RooAbsData} \\
\hline
\end{tabular}
\end{center}
\caption{Correspondence between mathematical concepts and {\tt RooFit}
classes.}
\label{tab:map}
\end{table}

Composite objects are built by creating all their components first.
For example, a Gaussian probability density function with its
variables

\begin{center}
\begin{tabular}{c}
$

G(x,m,s) = \frac{\exp \left( {-\frac{1}{2}\left( \frac{x-m}{s} \right)^2} \right)}
             {\int_{x_L}^{x_H} \exp \left( {-\frac{1}{2} \left( \frac{x-m}{s} \right)^2} \right)}
$
\end{tabular}
\end{center}

\noindent is created as follows:
\begin{center}
\begin{tabular}{c}
{\small \verb+  RooRealVar x("x","x",-10,10) ;            + } \\
{\small \verb+  RooRealVar m("m","mean",0) ;              + } \\
{\small \verb+  RooRealVar s("s","sigma",3) ;             + } \\
{\small \verb+  RooGaussian g("g","gauss(x,m,s)",x,m,s) ; + } \\
\end{tabular}
\end{center}
Each object has a name, the first argument, and a title, the second
argument.  The name serves as unique identifier of each object, the
title can hold a more elaborate description of each object and only
serves documentation purposes. All objects can be inspected with the
universally implemented {\tt Print()} method, which supports three
verbosity levels. In its default terse mode, output is limited to one
line, e.g.

\begin{center}
\begin{small}
\begin{tabular}{l}
{\verb+ root> f.Print() ;                          +} \\
{\verb@ RooRealVar::f:  0.531 +/- 0.072 L(0 - 1)@}\\
\end{tabular}
\end{small}
\end{center}

Object that represent variables, such as {\tt RooRealVar} in the
example above, store in addition to the value of that variable a
series of associated properties, such as the validity range, a
binning specification and their role in fits (constant
vs. floating), which serve as default values in many situations.

Function objects are linked to their ingredients: the function object
{\ttfamily g} {\em always} reflects the values of its input variables
{\ttfamily x,m} and {\ttfamily s}. The absence of any explicit
invocation of calculation methods allows for true symbolic
manipulation in mathematical style.

{\tt RooFit} implements its data models in terms of probability
density functions, which are by definition positive definite and unit
normalized:
\begin{equation}
\int_{\vec{x}_{min}}^{\vec{x}_{max}} f(\vec{x}) d\vec{x}~ \equiv 1, ~~~ F(\vec{x},\vec{p}) \ge 0 
\end{equation}
\noindent One of the main benefits of probability density functions over regular
functions is increased modularity: the interpretation of PDF
parameters is independent of the context in which the PDF is used.

The normalization of probability density functions, traditionally one
of the most difficult aspects to implement, is handled internally by
{\tt RooFit}: all PDF objects are automatically normalized to unity.
If a specific PDF class doesn't provide its normalization internally,
a variety of numerical techniques is used to calculate the normalization.

Composition of complex models from elementary PDFs is straightforward:
a sum of two PDFs is a PDF, the product of two PDFs is a PDF. The {\tt
RooFit} toolkit provides as set of 'operator' PDF classes that
represent the sum of any number of PDFs, the product of any number of
PDFs and the convolution of two PDFs. 

Existing PDF building blocks can be tailored using standard
mathematical techniques: for example substituting a variable
with a function in the preceding code fragment,
\begin{eqnarray*}
\nonumber m \rightarrow & m(m_0,m_1,y)& = m_0 + y*m_1 \\
\nonumber  & \downarrow & \\
\nonumber  G(x,&m(m_0,m_1,y)&,s) = G(x,y,m_0,m_1,s) \\
\end{eqnarray*}
\noindent is represented in exactly the same style in {\tt RooFit} code:
\begin{center}
\begin{tabular}{c}
{\small \verb+  RooRealVar m0("m0","mean offset",0) ;                + }\\
{\small \verb+  RooRealVar m1("m1","mean slope",1) ;                 + }\\
{\small \verb+  RooRealVar y("y","y",0,3) ;                          + }\\
{\small \verb@  RooFormulaVar m("m","m0+y*m1",                       @ }\\
{\small \verb@                  RooArgList(m0,m1,y)) ;               @ }\\
{\small \verb+  RooGaussian g("g","gauss(x,m,s)",x,m,s) ;            + }\\
\end{tabular}
\end{center}
\noindent Free-form interpreted C++
function and PDF objects such {\ttfamily RooFormulaVar} in the example
above, are available to glue together larger building blocks. The
universally applicable function composition operators and free-style
interpreted functions make it possible to write probability density
functions of arbitrary complexity in a straightforward mathematical
form.

\section{Composing and Using Data Models}

% slide 8 - (Re)using standard components
% slide 9 - (Re)using standard components
% slide 10 - (Re)using standard components
% slide 11 - (Re)using standard components
% slide 12 - (Re)using standard components

% slide 13 - Overview
% slide 14 - Fitting options
% slide 16 - Toy MC Generation
% slide 17 - Plotting
% slide 18 - Plotting
% slide 19 - Plotting

We illustrate the process of building a model and its various uses
with some example cases.

\subsection{A One-Dimensional Yield Fit}

The simplest and most common use of {\tt RooFit} is to combine two or
more library PDFs into a composite PDF to to determine the yield of
signal and background events in a one-dimensional dataset.

The {\tt RooFit} models library provides more than 20 basic
probability density functions that are commonly used in high energy
physics applications, including basics PDFs such Gaussian,
exponential and polynomial shapes, physics inspired PDFs, e.g. decay
functions, Breit-Wigner, Voigtian, ARGUS shape, Crystal Ball shape,
and non-parametric PDFs (histogram and KEYS\cite{KEYS}). 

In the example below we use two such PDFs: a Gaussian and an ARGUS
background function:

\begin{center}
\begin{small}
\begin{tabular}{l}
{\verb+ // Observable +} \\
{\verb+ RooRealVar mes("mes","mass_ES",-10,10) ;+ } \\
{\verb+ +} \\
{\verb+ // Signal model and parameters +} \\
{\verb+ RooRealVar mB("mB","m(B0)",0) ;+ } \\
{\verb+ RooRealVar w("w","Width of m(B0)",3) ;+ } \\
{\verb+ RooGaussian G("G","G(meas,mB,width)",mes,mB,w) ; + } \\
{\verb+ +} \\
{\verb+ // Background model and parameters +} \\
{\verb+ RooRealVar m0("m0","Beam energy / 2",-10,10) ;+ } \\
{\verb+ RooRealVar k("k","ARGUS slope parameter",3) ;+ } \\
{\verb+ RooArgusBG A("A","A(mes,m0,k)",mes,m0,k) ; + } \\
{\verb+ +} \\
%\end{tabular}
%\begin{tabular}{||l}
% \cline{1-1}
{\verb+ // Composite model and parameter+}\\
{\verb+ RooRealVar f("f","signal fraction",0,1) ;+} \\
{\verb@ RooAddPdf M("M","G+A",RooArgList(g,a),f) ;@} \\
% \cline{1-1}
\end{tabular}
\end{small}
\end{center}

\noindent The {\tt RooAddPdf} operator class {\tt M} combines the signal
and background component PDFs with two parameters each into a
composite PDF with five parameters:
\begin{eqnarray}
\nonumber M(m_{ES};m_B,w,m_0,k,f)~=~     f & \cdot & G(m_{ES};w,g)  \\
\nonumber                         + ~(1-f) & \cdot & A(m_{ES};m_0,k).
\end{eqnarray}

\noindent Once the model {\tt M} is constructed, a maximum likelihood
fit can be performed with a single function call:

\begin{center}
\begin{small}
\begin{tabular}{l}
{\verb+ M.fitTo(*data) ;         +} \\
\end{tabular}
\end{small}
\end{center}

\noindent Fits performed this way can be unbinned, binned and/or weighted,
depending on the type of dataset provided\footnote{Binned data can be
imported from a ROOT {\tt TH1/2/3} class, unbinned data can be
imported from a ROOT {\tt TTree} or a ASCII data file.}. The result of
the fit, the new parameter values and their errors, are immediately
reflected in the {\tt RooRealVar} objects that represent the
parameters of the PDF, {\tt mB,w,m0,k} and {\tt f}. Parameters can be
fixed in a fit or bounded by modifying attributes of the parameter
objects prior to the fit:

\begin{center}
\begin{small}
\begin{tabular}{l}
{\verb+ m0.setConstant(kTRUE) ;                    +} \\
{\verb+ f.setFitRange(0.5,0.9) ;        +} \\
\end{tabular}
\end{small}
\end{center}

Visualization of the fit result is equally straightforward:

\begin{center}
\begin{small}
\begin{tabular}{l}
{\verb+ RooPlot* frame = mes.frame() ;             +} \\
{\verb+ data->plotOn(frame) ;+}\\
{\verb+ M.plotOn(frame) ;+}\\
{\verb+ M.plotOn(frame,Components("A"),        +} \\
{\verb+                LineStyle(kDashed)) ;+}\\
{\verb+ frame->Draw()+}\\
\end{tabular}
\end{small}
\end{center}

\begin{figure}[hbt]
\begin{center}
\includegraphics[width=80mm]{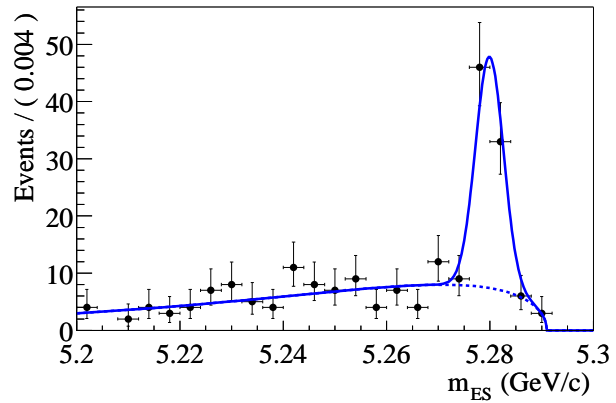}
\caption{One dimensional plot with histogram of a dataset,
overlaid by a projection of the PDF {\tt M}. The histogram error are
asymmetric, reflecting the Poisson confidence interval corresponding
to a $1 \sigma$ deviation. The PDF projection curve is automatically
scaled to the size of the plotted dataset. The points that define the
curve are chosen with an adaptive resolution-based technique that
ensures a smooth appearance regardless of the dataset binning.}
\label{fig:ex1}
\end{center}
\end{figure}

Figure \ref{fig:ex1} shows the result of the {\tt frame->Draw()}
operation in the above code fragment. A {\tt RooPlot} object
represents a one-dimensional view of a given observable. Attributes of
the {\tt RooRealVar} object {\tt mes} provide default values for the
properties of this view (range, binning, axis labels,...).

The {\tt plotOn()} methods of datasets and functions accept optional
arguments that modify the style and contents of what is drawn. The
second {\tt M.plotOn()} call in the preceding example illustrates some
of the possibilities for functions: only the {\tt A} component of the
composite model {\tt M} is drawn and the line style is changed to a
dashed style. Additional options exists to modify the line color,
width, filling, range, normalization and projection technique.  The
curve of the PDF is automatically normalized to the number of events
of the dataset last plotted in the same frame. The points of the curve
are chosen by an adaptive resolution-based technique: the deviation
between the function value and the curve will not exceed a given
tolerance\footnote{ Tolerance is preset at 0.1\% of the plot scale and
recursively evaluated halfway between each adjacent pair of curve
points} regardless of the binning of the plotted dataset.

The {\tt plotOn()} method of datasets accepts options to change the
binning, including non-uniform binning specifications, error
calculation method and appearance. The default error bars drawn for a
dataset are asymmetric and correspond to a Poisson confidence interval
equivalent to $1 \sigma$ for each bin content. A sum-of-weights error
($\sqrt{\Sigma_i w_i^2}$) can optionally be selected for use with
weighted datasets. A special option, {\tt Asym()}, is available to show
asymmetry distributions of the type $\frac{N_A - N_B}{N_A + N_B}$.
The errors bars will reflect a binomial confidence interval for such
histograms.

\subsection{A Simple Monte Carlo Study}

{\tt RooFit} PDFs universally support parameterized Monte Carlo event
generation, e.g.

\begin{center}
\begin{small}
\begin{tabular}{l}
{\verb+ RooDataSet* mcdata = M.generate(mes,10000) ;         +} \\
\end{tabular}
\end{small}
\end{center}

\noindent generates 10000 events in {\tt mes} with the distribution
of model {\tt M}. 

Events are by default generated with an accept/reject sampling
method. PDF classes that are capable of generating events in a more
efficient way, for example {\tt RooGaussian}, can advertise and
implement an internal generator method that will be used by {\tt
generate()} instead. Composite PDFs constructed with {\tt RooAddPdf}
delegate event generation to the component event generators for
computational efficiency.

\begin{figure}[hbt]
\begin{center}
\includegraphics[width=80mm]{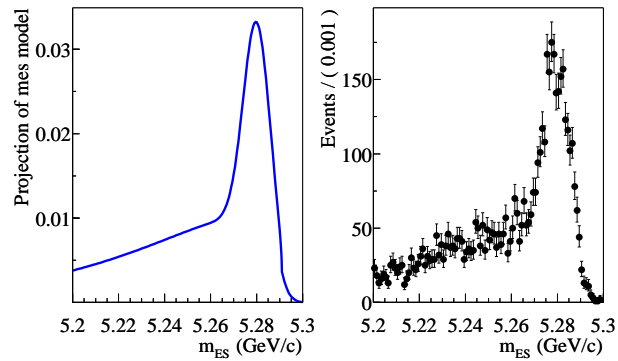}
\label{fig:ex2}
\caption{{\em Left:} Shape of PDF {\tt M}. {\em Right:} Distribution of 10000 
events generated from PDF {\tt M}}
\end{center}
\end{figure}

A common use of parameterized Monte Carlo is to study the stability and
bias of a fit, in particular when when statistics are
low\footnote{(Likelihood) fits can exhibit an intrinsic bias that
scale like $1/N$, where $N$ is the number of events in the fitted
dataset. At high statistics this bias is usually negligible compared
to the statistical error, which scales like $1/\sqrt{N}$, but at low
$N$ the effect may be significant. See e.g. Eadie et al\cite{eadie}
for details.}. A special tool is provided that automates the fitting
and generating cycle for such studies and collects the relevant statistics.
The example below examines the bias in the fraction parameter {\tt fsig}
of model {\tt M}:

\begin{center}
\begin{small}
\begin{tabular}{l}
{\verb+ // Generate and fit 1000 samples of 100 events+} \\
{\verb+ RooMCStudy mgr(M,M,mes) ;         +} \\
{\verb+ mgr.generateAndFit(100,1000) ; +} \\
{\verb+ +} \\
{\verb+ // Show distribution of fitted value of 'fsig' +} \\
{\verb+ RooPlot* frame1 = mgr.plotParam(fsig) ; +} \\
{\verb+ +} \\
{\verb+ // Show pull distribution for 'fsig' +} \\
{\verb+ RooPlot* frame2 = mgr.plotPull(fsig) ; +} \\
\end{tabular}
\end{small}
\end{center}

\begin{figure}[hbt]
\begin{center}
\includegraphics[width=80mm]{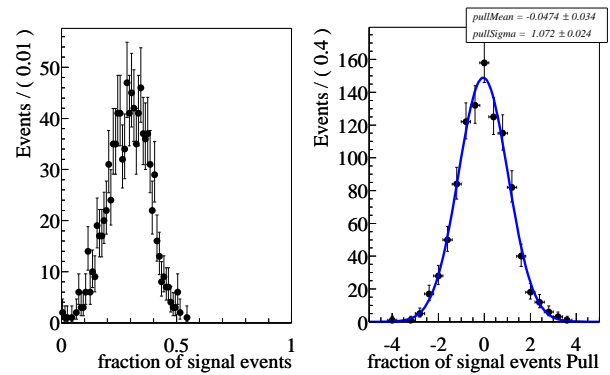}
\label{fig:ex3}
\caption{{\em Left:} distribution of fitted value of parameter {\tt f}
         of model {\tt M} to 1000 Monte Carlo data sets
         of 100 events each. {\em Right:} Corresponding pull distribution}
\end{center}
\end{figure}

\subsection{Multi-Dimensional Models}

Multi-dimensional data models are a natural extension of
one-dimensional models in data analysis. Use of additional observables
in the model enhances the statistical sensitivity of a fit, but are
traditionally less frequently used due to additional logistical and
computational challenges. While {\em fitting} a multi-dimensional
dataset is no more complex than fitting a one-dimensional dataset, the
variety of ways in which the data and model can be visualized is much
larger. In addition, the projection of a multi-dimensional model on a
lower-dimensional view often requires non-trivial computations. {\tt
RooFit} greatly automates the visualization process and other aspects
of multi-dimensional models so that use of multi-dimensional models is
not substantially more complicated that use of one-dimensional models.

Multi-dimensional models can be constructed in a variety of ways:

\begin{itemize}
   \item as a product of 1-dimensional PDFs
   \item as a fundamental multi-dimensional PDF object
   \item as a modified PDF that was originally intended for 1-dimensional use, e.g. \\
         $G(x;m,s) \to G(x;m(y,a,b),s) = G(x,y;a,b,s)$
\end{itemize}

We will illustrate some of the visualization options with a simple
3-dimensional model constructed as the product of three Gaussians
summed with a product of three polynomials:

\begin{displaymath}
  M = f \cdot G(x) G(y) G(z) + (1-f) \cdot P(x) P(y) P(z)
\end{displaymath}

\noindent encoded in {\tt RooFit} classes as follows

\begin{center}
\begin{small}
\begin{tabular}{l}
{\verb+  RooRealVar x("x","x",-10,10) ;               +  } \\
{\verb+  RooRealVar mx("mx","mean x",0) ;             + } \\
{\verb+  RooRealVar sx("sx","sigma x",3) ;            + } \\
{\verb+  RooGaussian GX("GX","gauss(x,mx,sx)",x,mx,sx);+ } \\
{\verb+  // Similar declarations of y,z, GY and GZ+ } \\
{\verb+  + } \\
{\verb+  RooRealVar ax("ax","bkg slope x",5) ;+ } \\
{\verb+  RooPolynomial PX("PX","PX(x,ax)",x,ax) ;+ } \\
{\verb+  // Similar declarations of PY and PZ+ } \\
{\verb+  + } \\
{\verb+  // Construct signal and background products+ } \\
{\verb+  RooProdPdf S("S","sig",RooArgSet(GX,GY,GZ)) ;+ } \\
{\verb+  RooProdPdf B("S","bkg",RooArgSet(PX,PY,PZ)) ;+ } \\
{\verb+  + } \\
{\verb+  // Construct sum of signal and background+ } \\
{\verb+  RooRealVar fsig("fsig","signal frac.",0.,1.) ;+ } \\
{\verb@  RooAddPdf M3("M3","S+B",RooArgList(S,B),fsig) ;@ } \\
{\verb+  + } \\
\end{tabular}
\end{small}
\end{center}

The {\tt RooProdPdf} class represents the product of two or more PDFs.
In most analysis applications such products factorize, e.g. $f(x)
\cdot g(y) \cdot h(z)$, but non-factorizing products or partially factorizing
products such as $f(x,y) \cdot g(y) \cdot h(z)$ are supported as well.
Factorization of multi-dimensional PDFs greatly simplifies many 
expressions involved in fitting, plotting and event generation,
and optimizations that take advantage of (partial) factorization
properties are automatically applied.

\subsubsection{Fitting}
The procedure to fit the 3-dimensional model {\tt M3} to a
3-dimensional dataset {\tt data3} is identical to that of the
1-dimensional model {\tt M}:

\begin{center}
\begin{small}
\begin{tabular}{l}
{\verb+ M3.fitTo(*data3) ;         +} \\
\end{tabular}
\end{small}
\end{center}

\subsubsection{Plotting}

A three-dimensional model like {\tt M3} can be visualized in a number
of ways. First there are the straightforward projections on the {\tt
x,y} and {\tt z} axes:

\begin{center}
\begin{small}
\begin{tabular}{l}
{\verb+ RooPlot* xframe = x.frame() ;             +} \\
{\verb+ data->plotOn(xframe) ;+}\\
{\verb+ model.plotOn(xframe) ;+}\\
{\verb+  + } \\
{\verb+ RooPlot* yframe = y.frame() ;             +} \\
{\verb+ data->plotOn(yframe) ;+}\\
{\verb+ model.plotOn(yframe) ;+}\\
{\verb+  + } \\
{\verb+ RooPlot* zframe = z.frame() ;             +} \\
{\verb+ data->plotOn(zframe) ;+}\\
{\verb+ model.plotOn(zframe) ;+}\\
\end{tabular}
\end{small}
\end{center}

While the invocation of the {\tt plotOn()} method is identical to the
one-dimensional case, the correspondence between the drawn curve and
the PDF is more complicated: to match the projected distributions of
multi-dimensional datasets, {\tt plotOn()} must calculated a matching
projection of the PDF. For a projection of a PDF $F(x,\vec{y})$
superimposed over the distribution of $x$ of a dataset $D(x,\vec{y})$,
this transformation is:

\begin{equation}
{\cal P}_{F}(x) = \frac{\int F(x,\vec{y}) d\vec{y}}{\int F(x,\vec{y}) dx d\vec{y}}
\label{eq:proj}
\end{equation}

For any dataset/PDF combination the set $\vec{y}$ of observables to be
projected is automatically determined: each {\tt RooPlot} object keeps
track of the 'hidden' dimensions of each dataset and matches those
to the dimensions of the PDF.

If the PDF $F$ happens to be a factorizing product, like our signal
and background PDFs {\tt S} and {\tt B}, Eq. \ref{eq:proj} reduces
to

\begin{equation}
{\cal P}_{F}(x) = \frac{ F_x(x) \int F_{\vec{y}}(\vec{y}) d\vec{y}}{\int F_x(x) dx \int F_{\vec{y}}(\vec{y}) d\vec{y}} = \frac{F_x(x)}{\int F_x(x) dx}
\end{equation}

This particular optimization is automatically recognized and
implemented by the {\tt RooProdPdf} components of the {\tt M3} model.
Non-factorizing multi-dimensional PDFs can also be drawn with {\tt
RooAbsPdf::plotOn()}, in such cases a combination of analytic and
numeric integration techniques is used to calculate the projection.

In addition to the full projections on the observables {\tt x,y} and
{\tt z}, it is often also desirable to view a projection of a {\em slice},
e.g. the projection on {\tt x} in a narrow band of {\tt y, z} or a box in {\tt
y} and {\tt z}. Such projections are often desirable because they
enhance the visibility of signal in the presence of a large
background:

\begin{center}
\begin{small}
\begin{tabular}{l}
{\verb+ // +\it Projection on X in a slice of Y} \\
{\verb+ RooPlot* xframe = x.frame() ;             +} \\
{\verb+ data->plotOn(xframe,"|y|<1") ;+}\\
{\verb@ model.plotOn(xframe,Slice(y,-1,+1)) ;@}\\
{\verb+  + } \\
{\verb+  + } \\
{\verb+ // +\it Projection on Z in a slice of X and Y} \\
{\verb+ RooPlot* zframe = z.frame() ;             +} \\
{\verb+ data->plotOn(zframe,"|x|<1&&|y|<1") ;+}\\
{\verb@ model.plotOn(zframe,Slice(x,-1,+1)@} \\
{\verb@                    ,Slice(y,-1,+1)) ;@}\\
\end{tabular}
\end{small}
\end{center}

\begin{figure}[hbt]
\includegraphics[width=80mm]{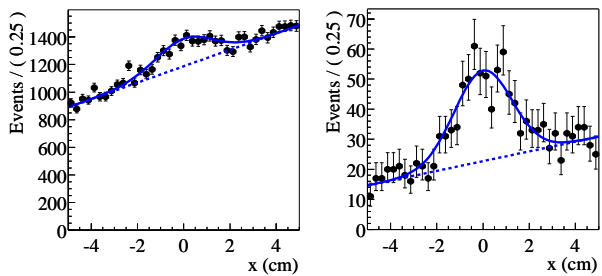}
\caption{Example of a likelihood projection plot of model {\tt M3}.
{\em Left:} projection of full dataset and PDF on {\tt x}. {\em
Right:} Projection of dataset and PDF with a cut on the likelihood
$L_{yz}$, calculated in the $(y,z)$ projection of the PDF, at -5.0. }
\label{fig:mcproj}
\end{figure}

While the {\tt Slice()} option implements a (hyper)cubic slice of the
data, {\tt RooFit} also supports Monte Carlo projection techniques
that allow to view regions of arbitrary shape. A common application of
this technique is the 'likelihood projection plot', where a
$n$-dimensional dataset and model are projected on one dimension after a
cut on the likelihood of the model in the remaining $n-1$
dimensions. Figure \ref{fig:mcproj} illustrates the power to enhance
the signal visibility of such a projection. The likelihood projection
technique naturally decomposes in a small number of {\tt RooFit}
operations: Figure \ref{fig:mcproj} has been created with less than 10
lines of macro code.

\subsubsection{Event Generation}

Event generation for multi-dimensional PDFs is similar
to that for one-dimensional PDFs. This example

\begin{center}
\begin{small}
\begin{tabular}{l}
{\verb+ RooDataSet* mcdata = +} \\
{\verb+          M3.generate(RooArgSet(x,y,z),10000) ;+} \\
\end{tabular}
\end{small}
\end{center}

generates a three-dimensional dataset of 10000 events.

If a model or its components are factorizable, such as the components
{\tt S} and {\tt B} of model {\tt M3}, the event generation will be
performed separately for each factorizing group of dimensions. There
is no limit to the number of observables in the event generator, but
the presence of groups of two, three or more of observables that do
not factorize will impact the performance of the accept/reject
technique due to phase-space considerations.

Sometimes it is advantageous to not let generate all observables of
the PDF, but take the distribution of some observables from an
external dataset. This technique is commonly applied in fit validation
studies where it is desirable to have a sample of Monte Carlo events
that replicates the statistical fluctuations of the data as precisely
as possible.  {\tt RooFit} facilitates this technique in the event
generator with the concept of 'prototype datasets'. The prototype
datasets prescribe the output of the event generator for a subset of
the observables to be generated. To use the prototype feature of the
generator, simply pass a dataset to the {\tt generate()} method. For
example, for a two-dimensional PDF {\tt M2}, this code

\begin{figure}[hbt]
\begin{center}
\includegraphics[width=80mm]{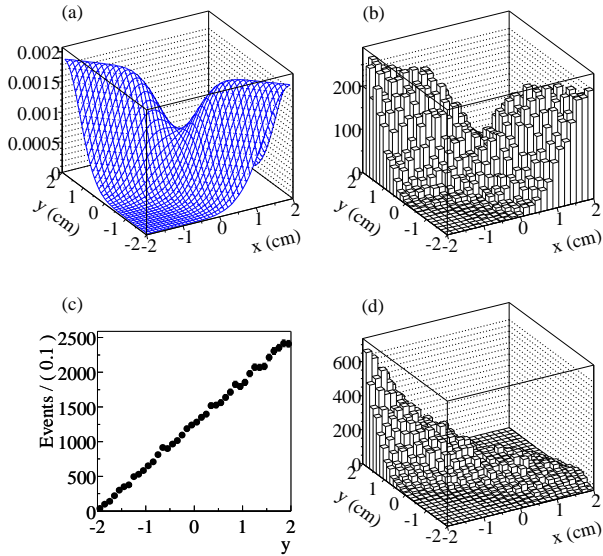}
\caption{Demonstration of prototype-based Monte Carlo event generation.
a) Two-dimensional PDF {\tt M2}.  b) Event sample generated from {\tt
M2}. c) One dimensional event sample in $y$ with linear distribution. d)
Event sample generated from {\tt M2} using event sample shown in c) as
prototype for $y$ distribution. }
\label{fig:ex4}
\end{center}
\end{figure}

\begin{center}
\begin{small}
\begin{tabular}{l}
{\verb+ RooDataSet* ydata ; +} \\
{\verb+ RooDataSet* mcdata = M2.generate(x,ydata) ;+} \\
\end{tabular}
\end{small}
\end{center}

generates a two-dimensional Monte Carlo event sample that has exactly
the same distribution in {\tt y} as the prototype dataset {\tt
ydata}, while the distribution of {\tt x} is generated such that it follows
the correlation between {\tt x} and {\tt y} that is defined by the PDF
{\tt M2}. Figure \ref{fig:ex4} illustrates the process:
Fig. \ref{fig:ex4}b shows the distribution of events generated the
regular way from PDF {\tt M2} (\ref{fig:ex4}a). Fig. \ref{fig:ex4}d
shows the distribution in {\tt (x,y)} when the distribution of events
in {\tt y} is taken from an external prototype dataset
(\ref{fig:ex4}c): the distribution of events in {\tt y} is exactly
that of the prototype while the correlation between {\tt x} and {\tt
y} encoded in the PDF is preserved.

\subsection{Advanced Fitting Options}

For fits that require more detailed control over the fitting process,
or fits that require non-standard extensions of the goodness-of-fit
quantity to be minimized, {\tt RooFit} provides an alternate interface
to the one-line {\tt fitTo()} method. In this procedure the creation
of the goodness-of-fit quantity to be minimized is separated from the
minimum finding procedure.

{\tt RooFit} provides two goodness-of-fit classes that represent the
most commonly used fit scenarios

\begin{itemize}
\item An (extended) negative log-likelihood implementation ({\tt RooNLLVar})
\item A $\chi^2$ implementation ({\tt RooChi2Var})
\end{itemize}

The $\chi^2$ implementation uses by default the asymmetric Poisson
errors for each bin of the dataset, but can be overridden to use
(symmetric) sum-of-weights error ($\sqrt{\Sigma_i w_i^2}$) instead,
for use with weighted datasets. Both classes represent their
goodness-of-fit variable as a regular {\tt RooFit} function so that
all standard techniques can be applied. For example, plotting the
likelihood defined by PDF {\tt M} and dataset {\tt data}
as function of parameter {\tt fsig} does not require any specialized
plotting methods:

\begin{center}
\begin{small}
\begin{tabular}{l}
{\verb+ RooNLLVar nll("nll","-log(L)",M,data) ; +} \\
{\verb+ RooPlot* frame = fsig.frame() ; +} \\
{\verb+ nll.plotOn(frame) ; +} \\
\end{tabular}
\end{small}
\end{center}

\begin{figure}[hbt]
\begin{center}
\includegraphics[width=80mm]{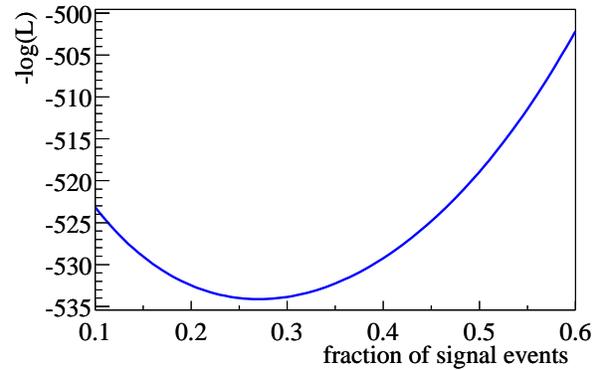}
\caption{Negative log-likelihood defined by PDF {\tt M}
and dataset {\tt data} as function of PDF parameter {\tt fsig}}
\label{fig:ex5}
\end{center}
\end{figure}

Figure \ref{fig:ex5} shows the plot that results from the
above code fragment.

The second step of the fit process, minimization of the
goodness-of-fit quantity, is performed with {\sc minuit}\cite{minuit},
via the interface class {\tt RooMinuit}:

\begin{center}
\begin{small}
\begin{tabular}{l}
{\verb+ // Initialize a minimization session +} \\
{\verb+ RooMinuit m(nll) ;                              +} \\
{\verb+ +} \\
{\verb+ // Invoke MIGRAD and HESSE +} \\
{\verb+ m.migrad() ;+} \\
{\verb+ m.hesse() ;+} \\
{\verb+ +} \\
{\verb+ // Change value and status of some parameters+} \\
{\verb+ p1.setConstant() ;+} \\
{\verb+ p2.setConstant() ;+} \\
{\verb+ p2.setVal(0.3) ;+} \\
{\verb+ +} \\
{\verb+ // Invoke MINOS +} \\
{\verb+ m.minos() ;+} \\
{\verb+ +} \\
{\verb+ // Save a snapshot of the fitter status+} \\
{\verb+ RooFitResult* r = m.save() ;+} \\
\end{tabular}
\end{small}
\end{center}

In the above example, the {\sc minuit} methods {\tt migrad} and {\tt
hesse} are invoked first. The effect of each {\sc minuit} operation is
immediately reflected in the {\tt RooRealVar} objects that represent
the parameters of the fit model. Conversely, any changes to the fit
parameter value or status are automatically propagated to {\sc
minuit}. The preceding code fragment illustrates how {\tt RooFit}
facilitates interactive fitting in C++. The {\tt RooMinuit}
interface almost completely isolates the user from any proprietary
syntax usually needed to interact with fitters like {\sc minuit}. One
only needs to be familiar with the meaning of the basic {\sc minuit}
procedures: {\tt migrad, hesse, minos, simplex, seek} and {\tt
contour}. The {\tt save()} method saves a full snapshot of the {\sc
minuit} state: the initial and current parameter values, the global
correlations, the full correlation matrix, status code and estimated
distance to minimum.

Since the {\tt RooMinuit} class can minimize {\em any} real-valued
{\tt RooFit} function, it is straightforward to fit customized
goodness-of-fit expression. For example one can add a
$\frac{1}{2}((p-\alpha)/\sigma_{\alpha})^2$ penalty term to a standard negative
log-likelihood using the {\tt RooFormulaVar} class:

\begin{center}
\begin{small}
\begin{tabular}{l}
{\verb+ RooNLLVar nll(nll,nll,pdf,data) ; +} \\
{\verb+ RooFormulaVar nllp("nllp","penalized nll",+}\\
{\verb@               "nll+0.5((p-alpha)/ealpha)^2",@} \\
{\verb@               RooArgList(nll,p,alpha,ealpha)) ;@} \\
\end{tabular}
\end{small}
\end{center}

Similar modifications can be made to $\chi^2$ definitions. It is also
possible to develop an entirely different goodness-of-fit quantity by
implementing a new class that derives from {\tt
RooAbsOptGoodnessOfFit}

\section{Efficiency and Optimal Function Calculation}

% slide 15 - Fitting speed and optimizations

As the complexity of fits increases, efficient use of computing
resources becomes increasingly important. To speed up the evaluation
of probability density functions, optimization techniques such as
value caching and factorized calculations can be used.

Traditionally such optimizations require a substantial programming
effort due to the large amount of bookkeeping involved, and often result
in incomplete use of available optimization techniques due to lack of
time or expertise. Ultimately such optimizations represent a
compromise between development cost, speed and flexibility.

{\tt RooFit} radically changes this equation as the object-oriented
structure of its PDFs allows centrally provided algorithms to analyze
any PDFs structure and to apply generic optimization techniques to it.
Examples of the various optimization techniques are

\begin{itemize}
\item {\em Precalculation of constant terms.} \\ In a fit, parts
 of a PDF may depend exclusively on constant parameters. These
 components can be precalculated once and used throughout the fit
 session.

\item { \em Caching and lazy evaluation}. \\ Functions are only
 recalculated if any of their input has changed. The actual
 calculation is deferred to the moment that the function 
 value is requested.

\item { \em Factorization.} \\ Objects representing a sum, product or 
convolution of other PDFs, can often be factorized from a single
N-dimensional problem to a product of N easier-to-solve 1-dimensional
problems.

\item { \em Parallelization.} \\ Calculation of likelihoods and other
goodness-of-fit quantities can, due to their repetitive nature, easily
be partitioned in to set of partial results that can be combined a
posteriori. {\tt RooFit} automates this process and can calculate partial
results in separate processes, exploiting all available CPU power
on multi-CPU hosts.

\end{itemize}

Optimizations are performed automatically prior to each potentially
CPU intensive operation, and are tuned to the specific operation that
follows. This realizes the maximum available optimization potential
for every operation at no cost for the user.

\section{Data and Project Management Tools}

As analysis projects grow in complexity, users are often confronted
with an increasing amount of logistical issues and bookkeeping tasks
that may ultimately limit the complexity of their analysis. {\tt
RooFit} provides a variety of tools to ease the creation and management
of large numbers of datasets and probability density functions
such as 

\begin{itemize}

\item {\em Discrete variables.} \\
A discrete variable in {\tt RooFit} is a variable with a finite set of
named states. The naming of states, instead of enumerating them,
facilitates symbolic notation and manipulation. Discrete variables can
be used to consolidate multiple datasets into a single dataset, where
the discrete variables states label the subset to which each events belongs.

\item {\em Automated PDF building.} \\ A common analysis technique is
to classify the events of a dataset $D$ into subsets $D_i$, and
simultaneously fit a set of PDFs $P_i(\vec{x},\vec{p}_i)$ to these
subsets $D_i$.  In cases were individually adjusted PDFs
$P_i(\vec{x},\vec{p}_i)$ can describe the data better than a single
global PDF $P(\vec{x},\vec{p})$, a better statistical sensitivity can
be obtained in the fit. Often, such PDFs do not differ in structure,
just in the value of their parameters. {\tt RooFit} offers a utility
class to automate the creation the the PDFs $P_i(\vec{x},\vec{p}_i)$:
given a prototype PDF $P(\vec{x},\vec{p})$ and a set of rules that
explain how the prototype should be altered for use in each subset
(e.g. 'Each subset should have its own copy of parameter {\tt foo}')
this utility builds entire set of PDFs $P_i(\vec{x},\vec{p}_i)$. It
can handle an unlimited set of prototype PDFs, specialization rules
and data subdivisions.

\item {\em Project configuration management.} \\
Advanced data analysis projects often need to store and retrieve
projection configuration, such as initial parameters values, names of
input files and other parameter that control the flow of
execution. {\tt RooFit} provides tools to store such information in a
standardized way in easy-to-read ASCII files. 
\end{itemize}

The use of standardized project management tools promotes structural
similarity between analyses and increases users abilities to
understand other {\tt RooFit} projects and to exchange ideas and code.

\section{Development Trajectory and Status}

% slide 21 - Development and use of RooFit in BaBar

{\tt RooFit} was initially released as {\ttfamily RooFitTools} in 1999
in the BaBar collaboration and started with a small subset of its
present functionality.  Extensive stress testing of the initial design
by BaBar physicists in 1999 and 2000 revealed a great desire to have a
tool like {\tt RooFit}, but identified a number of bottlenecks and
weaknesses, that could only be mitigated by a complete redesign.

The redesign effort was started in late 2000 and has introduced many
of the core features that define {\tt RooFit} in its current form:
strong mathematical styling of the user interface, nearly complete
absence of any implementation-level restrictions of any PDF building
or utilization methods, efficient automated function optimization and
powerful data and PDF management tools.

The new package, renamed {\ttfamily {\tt RooFit}}, has been available
to BaBar users since fall 2001. The functionality has been quite
stable since early 2002 and most recent efforts have been spent on
stabilization, fine tuning of the user interface and documentation.
At present five tutorials comprising more than 250 slides and 20
educational macros are available, as well as a reference manual
detailing the interface of all {\tt RooFit} classes.

Since October 2002 {\tt RooFit} is available to the entire HEP
community: the code and documentation repository has been moved from
BaBar to SourceForge, an OpenSource development platform, which
provides easy and equal access to all HEP users.
({\ttfamily http://roofit.sourceforge.net}),

\section{Current Use and Prospects}

Since the package's initial release {\tt RooFit} has been adopted by
most BaBar physics analyses. Analysis topics include searches for rare
B decays, measurements of B branching fractions and CP-violating rate
asymmetries, time-dependent analyses of B and D decays to measure
lifetime, mixing, and symmetry properties, and Dalitz analyses of B
decays to determine form factors.

The enthusiastic adoption of {\ttfamily RooFit} within BaBar
demonstrates the clear need and benefits of such tools in particle
physics. Since its migration to SourceForge, {\tt RooFit} is steadily
gaining users from other HEP collaborations.
\vspace{0.6cm} \\

%On a longer time scale
%{\tt RooFit} may be integrated in the ROOT distribution.

%\begin{acknowledgments}
%\end{acknowledgments}



\begin{thebibliography}{9}   % Use for  1-9  references
%\begin{thebibliography}{99} % Use for 10-99 references


\bibitem{sourceforge} {\tt http://roofit.sourceforge.net}
\bibitem{paw} R. Brun et al., {\em Physics Analysis Workstation}, CERN Long Writeup Q121
\bibitem{root} R. Brun and F Rademakers, {\em ROOT - An Object Oriented Data Analysis Framework},
Proceedings AIHENP'96 Workshop, Lausanne, Sep. 1996, Nucl. Inst. \&
Meth. in Phys. Res. A 389 (1997) 81-86.  See also {\tt
http://root.cern.ch}
\bibitem{KEYS} K. Cranmer, {\em Kernel Estimation in High-Energy Physics}, Comp. Phys. Comm {\bf 136}, 198-207 (2001).
\bibitem{eadie} Eadie et al, Statistical Methods in Experimental Physics, North Holland Publishing (1971)
\bibitem{minuit} F. James, {\em MINUIT - Function Minimization and Error Analysis}, CERN Long Writeup D506

\end{thebibliography}
\end{document}